\documentclass{article}
\usepackage{amsmath,graphicx,mlspconf,booktabs}
\usepackage[bookmarks=false]{hyperref}

\toappear{2019 IEEE International Workshop on Machine Learning for Signal Processing, Oct.\ 13--16, 2019, Pittsburgh, PA, USA}

\title{TOWARDS ETHICAL CONTENT-BASED DETECTION OF ONLINE INFLUENCE CAMPAIGNS}
\name{Evan Crothers\textsuperscript{\normalfont 1} \qquad Nathalie Japkowicz \textsuperscript{\normalfont 1,2} \qquad Herna L Viktor\textsuperscript{\normalfont 1}}
\address{\textsuperscript{1}University of Ottawa, ON, Canada \qquad \textsuperscript{2}American University, Washington DC, USA}

\begin{document}
\ninept

\maketitle
\begin{abstract}

The detection of clandestine efforts to influence users in online communities is a challenging problem with significant active development.  We demonstrate that features derived from the text of user comments are useful for identifying suspect activity, but lead to increased erroneous identifications (false positive classifications) when keywords over-represented in past influence campaigns are present.  Drawing on research in native language identification (NLI), we use ``named entity masking" (NEM) to create sentence features robust to this shortcoming, while maintaining comparable classification accuracy.  We demonstrate that while NEM consistently reduces false positives when key named entities are mentioned, both masked and unmasked models exhibit increased false positive rates on English sentences by Russian native speakers, raising ethical considerations that should be addressed in future research.

\end{abstract}
\section{Introduction}
\label{sec:intro}

\subsection{Overview}
\label{ssec:overview}

The recent surge in media coverage of nation-state efforts to influence public perception in online communities, as well as the release of large datasets by major social media platforms such as Facebook \cite{schrage_ginsberg_2018}, Reddit \cite{redditsuspicious}, and Twitter \cite{twitterei}, has led to a notable increase in research into the detection of online influence campaigns.

In haste to develop countermeasures, with particular concern for the possibility of foreign interference in upcoming elections, research focused on separating genuine user accounts from accounts linked to online influence campaigns (hereafter referred to as ``influence accounts") has rapidly outpaced research into characterizing the behaviour of these classification models.  An understanding of the shortcomings of existing state-of-the-art methods is important to demonstrate the fairness and legitimacy of such models as a means of arbitrating online communities.  These concerns are particularly important as the automated suppression of speech presents significant ethically considerations.  As such, the development of robust techniques for evaluation and reduction of algorithmic bias must develop alongside new detection methods.

In the context of this paper, ``L1" refers to an individual's first (native) language, while ``L2" refers to an individual's second (non-native) language.  Historically, the largest reported online influence campaigns targeting English users have been operated by countries with populations that do not typically speak L1 English.  As a result of this, content-based natural language processing (NLP) models trained on text from past influence campaigns may inadvertently develop a significant bias towards detection of writing by L2 English speakers --- particularly those who share an L1 language with the country to which the influence campaign has been attributed.

\subsection{Reddit}
\label{ssec:reddit}

Reddit is a news aggregation and discussion website that has dramatically risen in popularity over the last several years.  At the time of writing it is currently the fifth most popular website in both Canada and the U.S. according to Alexa rankings \cite{alexaus}\cite{alexaca}, placing it ahead of Twitter, Instagram, and Wikipedia.  Reddit allows for comments on submitted posts, and allows users to reply and vote to the comments of other users.  This results in a volume of discussion not found on other popular social media platforms.  Comments on Reddit are limited to 10,000 characters, allowing for much more verbose discussion than is easily possible on Twitter, which officially doubled its maximum character length to 280 in November, 2017 \cite{twitterlength}.

Reddit enables users to create communities around specific topics with few restrictions.  Concern has been expressed by Reddit users and moderators within that the site may be the target of ongoing nation-state efforts to influence popular opinion in order to support political goals.  Reddit CEO Steve Huffman addressed the community during the website's 2017 Transparency Report to address these concerns and report the staff's findings \cite{RedditTransparencyReport}.  This report included a release of 944 accounts ``of suspected Russian Internet Research Agency origin".  The sentences within comments from these 944 suspect accounts are considered the positive case for our classification task.

This paper is limited to open-source data that can be leveraged by independent researchers without special access to internal company data.  Open-source methods are valuable as they increase the number of individuals who can scrutinize activity in online spaces, exposing interference by coordinated groups, without requiring a privileged relationship with the platform holder.  With this in mind, however, the most effective techniques will likely require access to the platform's internal data.

\subsection{Ethical Focus}
\label{ssec:ethicalintro}

There is a distinction between authentic speech from those representing themselves online, and speech written under the direction of a government with the intent of manipulating a populace --- particularly when it is designed to misrepresent the author or spread misinformation.  The goal of detection systems should be to differentiate genuine expression from deliberate manipulation, focusing on signs that may indicate that online activity is disingenuous and directed.

It is not unexpected that a model trained to positively classify sentences written by Russia-operated influence accounts would demonstrate increased false positive rates (proportion of negative examples erroneously classified as positive) on English comments by L1 Russian speakers.  Similarly, it is reasonable that the false positive rate increases if the sentence contains named entities frequently found within the training data, such as those referring to American politics or cryptocurrency.  The combination of these tendencies, however, sets the groundwork for the automated suppression of speech --- and in particular, political speech --- by native Russian speakers.  This represents a serious ethical consideration that should influence the decision to deploy any content-based influence campaign detection model.  Influence campaigns operated by other countries with a high proportion of non-native English speakers will likely cause other language populations to face a similar risk.

User-submitted comments provide a significant variety of features that are useful for classification problems.  Features derived from the textual content of the comment itself --- or content-based features --- have been shown to hold predictive power on a number of classification problems related to the writer of the text \cite{stylometry}, including work specifically on detection of online influence campaigns \cite{redditbert}.  It is unrealistic to expect that future development in influence campaign detection should ignore a rich set of feature data.  Development of content-based models is encouraged to continue with careful consideration to potential algorithmic bias against language communities at high-risk of false positive classification.  Such mitigations may include negative training examples from L2 English language communities, or ensemble metadata classifiers to differentiate between genuine accounts and influence accounts with otherwise similar content features.

\section{Related Work}
\label{sec:pastwork}

Past work during the 2017 NLI Shared Task \cite{malmasi-etal-2017-report} has explored the state-of-the-art in NLI, demonstrating successful combinations of semantic and syntactic features for differentiating language learners from native English speakers.  However, this task did not reflect highly fluent advanced non-native speakers, which represent a much more challenging classification task.  Further research identified that the language level of L2 English Reddit users posting in European communities was much more sophisticated than most English learners and approached the level of the majority-English Reddit community as a whole \cite{DBLP:journals/tacl/RabinovichTW18}\cite{Kyle_Crossley_2014}.  Classification of sophisticated non-native English speakers on Reddit was the subject of a comprehensive analysis that included both comment content and metadata \cite{DBLP:conf/emnlp/GoldinRW18}.

NLI research has also contributed the concept of ``topic bias" \cite{malmasi-etal-2017-report}\cite{Brooke2012MeasuringIN} as an undesirable property in NLI datasets.  Topic bias may occur when the key themes and topics of texts are not evenly distributed across classes.  Within online influence campaigns, there is a significant skew towards political topics within comments by influence accounts.  As a result, positive detection may be heavily influenced by the presence of these topics.  This creates weaknesses in the classification model as future influence campaigns may not refer to the same topics as past influence campaigns, and presence of discussion of specific topics may cause a classifier to perform well on randomly sampled data, but poorly on data with similar topic content.  Named entity masking (NEM) has been used as an effective means of reducing topic bias in past NLI work \cite{DBLP:journals/tacl/RabinovichTW18}\cite{malmasi-dras-2014-arabic}\cite{malmasi-2016-subdialectal}, and should be applied to diminish topic bias within content-based influence campaign detection as well.

Prior work on detection of influence campaigns has mentioned L2 language features, such as differing stopword frequencies \cite{DBLP:journals/corr/abs-1901-11162}.  Much of the more formal research on influence campaigns focuses around Twitter due to the substantial quantity of available data \cite{twitterei}.  Past work has demonstrated a holistic approach to troll detection designed to incorporate features intended to match propaganda agents as well \cite{Fornacciari2018}.  While Reddit has seen an enormous surge in popularity, little formal research has been performed so far on online influence campaigns on Reddit, with a handful of graduate research projects forming the current state-of-the-art for classification \cite{redditbert}\cite{commentprediction}.

The project which currently claims the highest classification accuracy on this problem \cite{redditbert}, leverages vector representations of sentences (sentence embeddings) created using recently released deep NLP model BERT \cite{DBLP:journals/corr/abs-1810-04805}.  BERT, which stands for ``bidirectional encoder representations from transformers", leverages the Transformer architecture and attention mechanism discussed in the paper ``Attention is All You Need" \cite{DBLP:journals/corr/VaswaniSPUJGKP17} and applies them to a language modelling task.  The result is a network that allows for the creation of fixed-length vectors that contain both forward and backward contextual information for every token in the input text.  When using BERT, an additional token, referred to as \texttt{[CLS]}, is inserted at the start of the sentence, and can be used to obtain a fixed-length vector representation of sentences of variable length.  BERT has demonstrated state-of-the-art performance on a number of sentence and sentence-pair classification tasks, and is utilized in this paper to demonstrate the relevance of the findings to current state-of-the-art models.

There is some similarity between the work in this paper and the field of forensic linguistics, which has seen useful applications for NLI in cybercrime investigations \cite{10.1093/police/pay097}.  A forensic attribution of an influence campaign to a particular nation based on linguistic data is beyond the scope of this paper, and would necessitate evaluating linguistic similarities between not just English and Russian, but other commonly used languages as well.

\section{Methodology}
\label{sec:methodology}

This paper focuses on analyzing the language characteristics of comments posted by accounts ``of suspected Russian Internet Research Agency origin" released by link-aggregation and discussion website Reddit within their 2017 transparency report on April 10, 2018 \cite{RedditTransparencyReport}.   Using state-of-the-art natural language processing (NLP) model BERT \cite{DBLP:journals/corr/abs-1810-04805}, contextual embeddings for sentences within these comments will be generated, and a classifier will be trained to distinguish between sentences from randomly sampled Reddit accounts and those from suspected influence accounts.  This classification methodology and training dataset is designed to be comparable to that used by the project that currently claims state-of-the-art classification performance on Reddit \cite{redditbert}, with the distinction of being a sentence-level, rather than comment-level, task. This process will be repeated with the same data after performing ``named entity masking" (NEM) to replace named entities with their corresponding parts-of-speech (POS) tag.

The performance of this model will be evaluated not only against a holdout test set of suspect sentences and random sentences (described in \S \ref{sssec:dataset1} and \S \ref{sssec:dataset2} respectively), but also against two separate evaluation datasets based on the L1 language of the user (described in \S \ref{sssec:dataset3}).  We create a dataset of sentences from comments by users who self-identify as being from L1 English countries, as well as a set of comments by users who self-identify as being from Russia.  These datasets are constructed using similar methodology to recent work in native language identification \cite{DBLP:conf/emnlp/GoldinRW18}.  This test is used to demonstrate the tendency of each model to generate more false positives when considering English comments written by users who speak Russian as a first language, as opposed to English native speakers.  Also evaluated is a more demanding test set which filters these sentences to those that contain ``frequent named entities" (FNE): the top ten named entities most frequently mentioned in suspect comments within the ground-truth data.

The purpose of this paper is to form a compelling case for the development of safeguards in the deployment of content-based moderation methods, particularly those that may target distinctive linguistic characteristics (e.g. L2 English) shared by users outside of the target group.  Online influence campaign detection offers a useful example where this problem is evident.  The results of this work offer some future direction for the leveraging of content-based features for influence account detection, which may be integrated into an ensemble model for influence campaign detection on Reddit, similar to recent work in building ensemble troll detection models on Twitter \cite{Fornacciari2018}.  Synthesizing these features with past work on ensemble approaches to influence campaign detection on Reddit \cite{commentprediction} may contribute to an ethically sound and effective approach.

\section{Experimental Setup}
\label{sec:experimentalsetup}

All experiments were conducted on a n1-standard-2 (2 vCPUs, 7.5 GB memory) Google Cloud instance, with a Tensor Processing Unit (TPU) v3-8. The code for the experiment is available online \cite{papercode}.

\subsection{Datasets}
\label{ssec:datasets}

This analysis relies on three corpora, as detailed below.

\subsubsection{Corpus I: Comments from 2017 Reddit transparency report}
\label{sssec:dataset1}

This dataset is comprised of Reddit comments made by accounts on a list released by Reddit staff on April 10, 2018 as ``of suspected Russian Internet Research Agency origin" \cite{redditsuspicious}\cite{RedditTransparencyReport}.  These accounts were preserved for the purposes of transparency, allowing users to scrape their comment histories for further analysis.  A full export of all comments made by these suspect accounts was performed by Alberto Coscia and is available on GitHub \cite{ALCC01}.

This corpus represents the only official collection of Reddit accounts released to date as ``suspicious" in the context of coordinated influence campaigns.  While other independent researchers have collected other lists of accounts which exhibit some suspicious behavior \cite{joshrussel}, the official designation of these accounts represents a stronger confidence level not found in other sources.  As such, for this analysis, we will only use the officially designated suspect accounts as ground-truth training examples.

As a platform holder, Reddit has access to additional data not publicly available.  This includes access to IP logs, private actions (such as upvotes/downvotes), and more granular user activity tracking.  These features, as well as Reddit's access and subject-matter expertise in their data, allows for this attribution to be considered accurate with a high degree of confidence.

\subsubsection{Corpus II: Randomly sampled Reddit comments}
\label{sssec:dataset2}

This dataset of random sampled comments was created by Brandon Punturo \cite{punturo}, and has been used in two past online influence detection projects on Reddit \cite{redditbert}\cite{commentprediction}.  The dataset represents a typical random sampling approach to acquiring a negative class for influence campaign detection.  We use this dataset to compare our results to past work in the field.  It is important to note that more sophisticated sampling techniques may better address false positive similarities between influence accounts and genuine accounts.

For the purpose of this study, we assume that none of these randomly-sampled comments are attached to an influence campaign, based on the current understanding of the scale of past influence campaigns \cite{redditsuspicious} and the volume of daily comments on Reddit \cite{pushshiftio}.

\subsubsection{Corpus III: Augmented L2 Reddit dataset}
\label{sssec:dataset3}

Reddit has been the data source for past work on Native-Language Identification (NLI) on sophisticated second-language speakers \cite{DBLP:journals/tacl/RabinovichTW18}\cite{DBLP:conf/emnlp/GoldinRW18}.  This work entailed the creation of datasets of Reddit comments from users of a variety of different languages by looking for self-identified ``flair" in European subreddits.  This dataset includes a sizeable number of comments from self-identified Russian users: 31,167 from European subreddits and 586,398 comments from other subreddits \cite{l2redditdataset}.

We augment the Russian content of this dataset by leveraging the user-specified flair in the subreddit ``AskARussian", which has 4,072 users at the time of writing.  We collect the comments of users with self-identified flairs that indicate Russia or a specific Russian region.  Deduplication is performed, and users who are already present in the original L2-Reddit dataset are discarded, as are self-declared bot accounts.  The results in a total of 774,702 comments.  When tokenized into sentences greater than 10 characters long, the result is 1.9 million sentences.  We believe this is the most comprehensive dataset of online comments made by highly-fluent L1 Russian / L2 English language speakers.  It is hoped this dataset will assist in future work into research of algorithmic bias.

The L2-Reddit dataset also includes comments by users who self-identify as being from countries that typically speak L1 English.  We assess Australia, Ireland, New Zealand, the United Kingdom, and the United States of America as fulfilling this criterion.  These L1 English comments will be compared to the L1 Russian comments to determine the difference in classification accuracy.

Similar to Corpus II, we assume that none of these comments are affiliated with online influence campaigns, based on the current understanding of the scale of past influence campaigns \cite{redditsuspicious} and the volume of daily comments on Reddit \cite{pushshiftio}.

\subsection{Data Preprocessing}
\label{sec:preproc}
  
The first two corpora are considered the two distinct classes for the classification task, and are used together to train a classifier.  Corpus III will be used to evaluate false positive rates of this classifier against native English and Russian speakers.

As this is a dataset composed of real comments in original formatting, data preprocessing is an important consideration.  The comment data is cleaned according to a multi-step process.

\begin{enumerate}
  \item Normalize datasets into similar formats using regular expressions to remove extraneous escape characters.
  \item Perform sentence tokenization using Python NLTK \cite{Loper02nltk:the} to extract sentences from comments.
  \item Remove newline characters, Reddit quote markdown characters, and horizontal tab characters (\texttt{\&\#009;}).
  \item Remove all URLs and replace with [URL] token.
  \item Discard sentences shorter than 10 characters.  Very small sentences are poorly-suited to the classifier and may introduce noise.
  \item Run full BERT tokenization pipeline \cite{bertgit}, which includes converting to lowercase, WordPiece \cite{wordpiece} tokenization, punctuation splitting, and invalid character removal.
\end{enumerate}

\subsection{BERT Sentence Embedding Classification}

Sentence embedding refers to the family of techniques whereby sentences are mapped to vector representations within a continuous vector space, as in the case of word and phrase embedding \cite{Bengio:2003:NPL:944919.944966}\cite{DBLP:journals/corr/MikolovSCCD13}.  These ``sentence embeddings" are useful for classification and clustering of sentences.  Fixed-length sentence embeddings are particularly valuable as they convert variable length sentences into fixed-length feature vectors.

Fixed-length sentence embeddings are obtained from BERT by feeding the WordPiece tokenized input into the BERT model and reading the final layer activations for the prepended special classification token (\texttt{CLS}), as per the intent of the model \cite{DBLP:journals/corr/abs-1810-04805}.  A simple configuration is used for BERT in order to emphasize repeatability, and avoid undue emphasis on the specifics of the neural network's construction.

A single layer classifier is trained on sentence embeddings from the BERT model, as is standard for fine-tuning BERT for sentence classification tasks.  A maximum sequence length of 128 is chosen for the model.  While BERT supports sequence lengths up to 512, a shorter sequence length is recommended by Google Research \cite{bertgit} as the relationship between Transformer attention and sequence length is quadratic \cite{DBLP:journals/corr/abs-1810-04805}, leading to dramatic increases in computation time.  A training batch size of 64 is used to maximize the efficiency of the TPU v3-8.  We use the uncased variant of the BERT base model (12-layer, 768-hidden, 12-heads, 110M parameters) for our experiment.

\subsection{Masking}
\label{sec:masking}

The available data on influence campaigns contains frequent mentions of specific named entities.  As a result, past classification work using BERT embeddings for classification on Reddit \cite{redditbert} has highlighted posts containing these words as a prominent failure case.  In order to demonstrate the impact that these keywords have on BERT feature classification, we perform named-entitiy recognition (NER) to extract named entities (NEs) from sentences in the three corpora described in \S \ref{ssec:datasets}.  These will be used to form three additional datasets, with each named entity replaced by the corresponding tag.  This approach, similar to that taken in native language identification research on Reddit \cite{DBLP:journals/tacl/RabinovichTW18}, emphasizes the other content features in the text, such as grammatical structure and word choice.  To perform named entity masking, we use the largest (and most accurate) implementation available in the ``spaCy" Python package, which supports detection of a broad range of entities at accuracy comparable to state-of-the-art \cite{spaCy}\cite{Kiperwasser2016SimpleAA}.

Table \ref{tab:ner-table-suspicious} shows the ten most frequent named entities in Corpus I, omitting less-distinctive results for ``DATE", ``CARDINAL", and ``PERCENT" entities.  We have also omitted one named ``PERSON" entity: ``\texttt{:D}".  While the frequent presence of this emoji within the suspicious comments may be a distinguishing feature, it does not meet the criteria of a valid named entity for this analysis.  Each unique named entity is only counted once per comment that it occurs in, to prevent highly repetitive comments in which a named entity is mentioned multiple times from dominating the results.

The entities in Table \ref{tab:ner-table-suspicious} will be considered ``frequent named entities" (FNEs) which we will used to filter the prepared datasets to create a final evaluation dataset that emphasizes known failure modes in content-based models.

\begin{table}
\centering
\begin{tabular}{@{}lllll@{}}
\toprule
Entity   & Entity Type & Count &  &  \\ \midrule
US      & GPE  & 79   &  &  \\
TIE      & ORG  & 79   &  &  \\
Trump    & ORG  & 67    &  &  \\
Bitcoin     & ORG  & 52    &  &  \\
Hillary      & PERSON & 39     &  &  \\
America      & GPE & 38     &  &  \\
Russia      & GPE & 37     &  &  \\
Russian     & NORP  & 31     &  &  \\ 
ISIS     & ORG  & 29     &  &  \\ 
BTC     & ORG  & 28     &  &  \\ \bottomrule
\end{tabular}
\vspace{0.3cm}
\caption{Most common named entities within Corpus I, used to generate additional ``frequent named entity" (FNE) evaluation dataset.} \label{tab:ner-table-suspicious}
\end{table}

\section{results}
\label{sec:results}

\begin{table}[ht]
\centering
\begin{tabular}{@{}lll@{}}
\toprule

& Unmasked Model & Masked Model \\ \midrule
Accuracy                                                           & \textbf{0.7409}   & 0.7266    \\
AUC                                                                & \textbf{0.7409}   & 0.7266    \\
F1 Score                                                           & \textbf{0.7433}   & 0.7302    \\
Precision                                                          & \textbf{0.7359}   & 0.7202    \\
Recall                                                             & \textbf{0.7512}   & 0.7409    \\ \bottomrule
\end{tabular}
\vspace{0.3cm}
\caption{Mean evaluation results on masked and unmasked models trained to differentiate between suspect sentences (positive class derived from Corpus I) and random comments (negative class derived from Corpus II).} \label{tab:test-results}
\end{table}

\begin{table}
\centering
\begin{tabular}{@{}lll|ll@{}}
\toprule
Dataset & Unmasked Model & Masked Model  & $t$-statistic \\ \midrule
L1Ru        & 63.82\%   & \textbf{43.72\%}      & 9.92          \\
L1En  &     38.82\%     &   \textbf{36.10\%}    & 3.67                  \\
L1Ru-FNE  & 70.09\%   & \textbf{56.55\%}        & 20.46          \\
L1En-FNE  &  54.46\%    &    \textbf{51.97\%}   & 3.80                   \\ \bottomrule
\end{tabular}
\vspace{0.3cm}
\caption{Type I error rates on Corpus III sentences written by L1 Russian and L1 English users, as well as $t$-statistic of performance difference.} \label{tab:l1ru-results}
\end{table}

Table \ref{tab:test-results} illustrates that the performance of the NE masked model (NEMM) is comparable to that of the unmasked model when distinguishing between sentences written by randomly sampled Reddit users and sentences written by suspected influence accounts.  The unmasked model does however retain a slight advantage on the trained classification task.

Table \ref{tab:l1ru-results} shows the performance of both models on an evaluation dataset of randomly sampled English-language comments by L1 Russian (L1Ru) and L1 English users (L1En).  Both models demonstrate a significant increase in false positives when applied to the L1Ru dataset compared to the L1En dataset.  The false positive rate is highest for the unmasked model when classifying comments by L1 Russian speakers that contain a named entity frequently mentioned within the training data (L1Ru-FNE), followed by the false positive rate on arbitrary L1 Russian comments (L1Ru).  The increased error caused by the presence of frequent named entities is substantially improved by the NEMM.

The results of each test in this experiment were rigorously validated by repeating 10 runs of 10-fold cross-validation (10x10 fold CV).  Each run was performed with a new set of random samples from the corpora.  We then perform significance testing by using the corrected repeated k-fold CV test to calculate the t-statistic \cite{Nadeau2003}\cite{10.1007/978-3-540-24775-3_3}.  For the accuracy of differentiating random sentences from suspect sentences as displayed in Table \ref{tab:test-results}, we attain a score of $t=4.9394$ (two-tailed $p<0.00001$).  Significance testing results for the difference in false positive rates are displayed alongside results in Table \ref{tab:l1ru-results}, computed using a paired-sample t-test.  All of the readings fall within a two-tailed significance level of $p < 0.001$.

\section{analysis}
\label{sec:analysis}

The results described in \S \ref{sec:results} indicate that models trained exclusively on content features of existing influence campaigns disproportionately misclassify speakers of that language, as well as users who use specific named entities common to past influence accounts.  When both of these conditions coincide, the effect is magnified substantially, giving the highest percentage of false positives in the evaluation set.

Simply put: users with Russian as a first language, particularly those who are discussing the United States, politics, or cryptocurrency, are at increased risk of false positive classification when writing in English.

When the classification model and test data is masked, the model becomes more resistant to the presence of FNEs and topic bias in L1 Russian comments, but a pronounced gap between the performance on L1 English and L1 Russian sentences remains.

\section{Acknowledgements}

Research supported with Cloud TPUs from Google's TensorFlow Research Cloud (TFRC).

\section{Conclusion}

We conclude that the use of content-based features without safeguards creates the potential for discrimination against users of specific language backgrounds, especially when they are engaged in speech that contains common named entities that often reflect political topics.  As protection of genuine free expression of political opinions on the Internet is a value of many organizations and governments, online influence detection models designed by social media platforms or government organizations should consider constructing test datasets of L2 English speakers using contextual data clues, such as flair or IP address, for the purpose of identifying potential avenues of discrimination.  While some measurable bias towards detection of users who speak the same L1 language as the target distribution is expected, this behaviour should be tracked and mitigated whenever possible.

\subsection{Future Work}

The evaluation of language classification models with regard to ethical considerations is a challenging and worthwhile area of research.  Broader analysis that evaluates a greater number of classifiers, and deeper research into mitigations against discrimination in the domain of influence campaign detection, would be a promising direction for future research.

The study of linguistic features that accurately target deceptive or manipulative behaviours may assist in addressing the ethical concerns highlighted in this paper.  A related field, the detection of hate speech inciting violence, is another worthwhile area of future development both in improving detection methods and in answering ethical questions around the classification boundary between personal opinion and hate speech.

Due to the observed linguistic similarities between English sentences by self-identified Russian users and sentences from influence accounts, our results may be interpreted in the context of attribution of influence campaigns to their originators.  Such an analysis is beyond the scope of this paper.  As the existing online influence campaign datasets released by Twitter \cite{twitterei} and Reddit \cite{redditsuspicious}\cite{ALCC01} do not contain any IP information that may be approached using geolocation techniques, it is difficult to independently determine the origin of suspect accounts.  A more thorough linguistic analysis similar to that performed by Goldin et al. \cite{DBLP:conf/emnlp/GoldinRW18} may provide more insight into online influence campaigns on this platform.  Furthermore, this could allow for detection of future trends, such as recruitment of native language speakers, or the use of generative text models, such as GPT-2 \cite{radford2019language}.

Detection of online influence campaigns is a field that is likely to change rapidly over the coming years as new advances in detection prompt corresponding advances in evasion.  Successful classifiers will likely require diverse content-based and metadata features to attain a solution that is both effective and ethical.

\bibliographystyle{ieeebib}
\bibliography{references.bib}

\end{document}